\documentstyle[preprint,psfig,aps]{revtex}

\begin{document}
\draft

\title{Application of time-dependent density functional
theory to optical activity}

\author{K. Yabana\footnote{E-mail: yabana@nt.sc.niigata-u.ac.jp}}
\address{Graduate School of Science and Technology, Niigata University\\
Niigata 950-21, Japan\\
and\\}
\author{G.F. Bertsch\footnote{E-mail: bertsch@phys.washington.edu}}
\address{Physics Department and Institute for Nuclear
Theory\\
University of Washington, Seattle, WA 98195 USA
}
\date{text/chiral/chiral.tex; Dec. 11, 1998}
\maketitle
\def\C76{C$_{76}$~}
\def\C60{C$_{60}$~}
\def\eq#1{eq. (\ref{#1})}
\def\pipi{$\pi-\pi^*$}
\def\sigsig{$\sigma -\sigma^*$}
\def\be{\begin{equation}}
\def\ee{\end{equation}} 

\begin{abstract}
As part of a general study of the time-dependent local density 
approximation (TDLDA), we here report calculations of optical activity 
of chiral molecules.
The theory automatically satisfies sum rules and the Kramers-Kronig 
relation between circular dichroism and optical rotatory power. 
We find that the theory describes the measured circular dichroism 
of the lowest states in methyloxirane with an accuracy of about 
a factor of two.  
In the chiral fullerene C$_{76}$ the TDLDA provides a consistent
description of the optical absorption spectrum, the circular
dichroism spectrum, and the optical rotatory power, except for
an overall shift of the theoretical spectrum.
\end{abstract}

\section{Introduction}

The time-dependent local density approximation (TDLDA) as well as
the time-dependent Hartree-Fock theory has been 
applied to the optical absorption of atomic and molecular systems 
with considerable success [1-14].  Here we want to see how well 
the TDLDA method does on a more subtle aspect of the optical response, 
the optical activity of chiral molecules.  
Calculation of circular dichroism and especially optical rotatory 
power is more challenging because the operators that must be
evaluated are more sensitive to the approximations on the wave 
function than the electric dipole in the usual form $e \vec r$.  
Nevertheless, we anticipate that TDLDA may be a useful theory
because the operators are still of the form of single-electron 
operators. The TDLDA is derived by optimizing a wave function 
constructed from (time-dependent) single-particle wave functions, 
so its domain of validity is the one-particle observables.

In our implementation of TDLDA \cite{ya96}, we represent the electron wave 
function on a uniform spatial grid. The real-time evolution of
the wave function is directly calculated and the response functions
are calculated by the time-frequency Fourier transformation.
The method respects sum rules and the Kramers-Kronig relation
between the circular dichroism and optical rotatory power.
Since the grid representation is bias-free with respect to 
electromagnetic gauge, it is not subject to the gauge difficulties 
encountered when the space of the wave function is constructed 
from an atomic orbital representation.

Optical activity has been a challenging problem for computational
chemistry, but there has been considerable progress in recent years.  
For example, Carnell et al. \cite{ca91} present a good description 
of the circular dichroism of excited states of R-methyloxirane 
using a standard Gaussian representation of the wave function.  
The optical rotatory power is a much more difficult observable, 
since the whole spectrum contributes. Only very recently have 
$ab~initio$ calculations been reported for this property\cite{po98,pe98}.

After presenting our calculational method, we report our exploratory
study on optical activities of two chiral molecules: 
R-methyloxirane, a simple 10-atom molecule 
with known chiroptical properties up to the first few excited 
states\cite{ca91,co83,br94}, and C$_{76}$, a fullerene with very 
large optical rotatory power and significant circular dichroism 
in the visible and UV\cite{ha93}.  

\section{Formalism}

\subsection{Some definitions}

Polarization of chiral molecule in applied electromagnetic field 
is expressed using two coefficients $\alpha$ and $\beta$ as \cite{co37}
\be
\vec p = \alpha \vec E 
- \frac{\beta}{c} \frac{\partial \vec H}{\partial t}.
\ee
Here $\alpha$ is the usual polarizability and is given microscopically as
\be
\alpha(E) = e^2 \sum_n \left(
\frac{1}{E_{n0} - E - i\delta} + \frac{1}{E_{n0} + E + i\delta}
\right)
\frac{1}{3} \langle \Phi_0 |\sum_i \vec r_i| \Phi_n \rangle^2,
\ee
where $\Phi_n$ and $E_n$ are the eigenvector and eigenvalue of 
the $n$-th eigenstate of the many-body Hamiltonian $H$,
$H\Phi_n = E_n \Phi_n$, and $E_{n0}=E_n-E_0$. The $\delta$ is an
infinitesimal positive quantity. Employing the oscillator
strength
\be
f_n = \frac{2mE_n}{\hbar^2} \frac{1}{3}
\langle \Phi_0 |\sum_i \vec r_i|\Phi_n \rangle^2,
\ee
we define the optical absorption strength whose integral is
normalized to the active electron number,
\be
S(E) = \sum_n \delta (E-E_n) f_n.
\ee
It is related to the imaginary part of the polarizability,
\be
S(E)=\frac{2mE}{\hbar^2 e^2} \frac{{\rm Im}\alpha(E)}{\pi}.
\ee

The basic quantity which characterizes the chiroptical transition
is the rotational strength defined by \cite{ha80}
\be
R_n = -\frac{e^2 \hbar}{2mc}
\langle \Phi_0 |\sum_i \vec r_i|\Phi_n \rangle \cdot 
\langle \Phi_n |\sum_i \vec r \times \vec \nabla |\Phi_0 \rangle.
\ee
We define the complex rotational strength function,
\be
{\cal R}(E) = \sum_n \left(
\frac{1}{E_{n0} - E - i\delta} - \frac{1}{E_{n0} + E + i\delta} \right) R_n.
\ee
The beta function in eq.(1) is related to ${\cal R}(E)$ by
$\beta (E) = \frac{\hbar c}{3E} {\cal R}(E)$.
We will also use the rotational strength function $R(E)$ defined by
\be
R(E) = \sum_n \delta(E-E_n)R_n = \frac{{\rm Im}{\cal R}(E)}{\pi}.
\ee
As is seen below, 
the optical rotatory power is proportional to the real part of $\beta(E)$,
and the circular dichroism to $R(E)$. They are related to each other
by the Moscowitz's generalized Kramers-Kronig relation \cite{mo62}.

The difference of complex indices of refraction for left and right 
circularly polarized light is proportional to ${\cal R}$ in dilute 
media; the relation is
\be
n_L-n_R = {8 \pi N_1\over 3 } {\cal R}(E),
\ee
where $N_1$ is the number of molecules per unit volume.  
For comparison with experiment, the common measure of circular 
dichroism is the decadic extinction coefficient, given by
\be
\Delta \epsilon = {4 \pi \over \lambda_{cm} C\log_e 10} {\rm Im} (n_L-n_R),
\ee
where $C$ is the concentration of molecules in moles/liter and 
the subscript on the wavelength $\lambda$ is a reminder to 
express it in centimeters.  The optical rotatory power is 
conventionally reported as
\be
[\alpha] = 180^\circ {10\over C_{gm}\lambda}{\rm Re} (n_L-n_R),
\ee
where $C_{gm}$ is the concentration of molecules in gm/cm$^3$.

\subsection{Real-time TDLDA}

We first rewrite the above strength functions as time integrations.
We employ the time dependent wave function 
$\Psi(t)=\exp[-iHt/\hbar]\Psi(0)$
with the initial wave function at $t=0$ given by
$\Psi(0)=\exp[ik\sum_i z_i]\Phi_0$,
where $k$ is a small wave number.
In the linear response, the time-dependent polarizability is proportional 
to the dipole matrix element,
$z(t)=\langle \Psi(t)|\sum_i z_i|\Psi(t) \rangle$.
The frequency dependent polarizability in $z$ direction is then obtained 
as the time-frequency Fourier transformation of $z(t)$,
\be
\alpha_z(E) = \frac{e^2}{k} \int_0^{\infty} \frac{dt}{\hbar}
e^{i(E+i\delta)t/\hbar} z(t).
\ee
The polarizability $\alpha(E)$ is given by the orientational average,
$\alpha = (\alpha_x+\alpha_y+\alpha_z)/3$.

Similarly, we denote the angular momentum expectation value as 
$ L_z(t) = \langle \Psi(t) | -i(\vec r \times \nabla)_z | \Psi(t) \rangle $. 
To linear order in $k$, we may express it as
\be
L_z(t) = -2k \sum_n \cos \left( \frac{E_{n0}t}{\hbar} \right)
\langle \Phi_0|\sum_i z_i|\Phi_n \rangle
\langle \Phi_n|\sum(\vec r_i \times \vec \nabla_i)_z|\Phi_0 \rangle.
\ee
The complex rotatory strength function ${\cal R}(E)$ is expressed as,
\begin{eqnarray}
{\cal R}_z(E) 
&=& \frac{e^2 \hbar}{2mc} \frac{i}{k}
\int_0^{\infty} \frac{dt}{\hbar} e^{(E+i\delta)t/\hbar} L_z(t) \nonumber \\
&=& \frac{e^2\hbar}{2mc} \sum_n \left(
\frac{1}{E_{n0}-E-i\delta} - \frac{1}{E_{n0}+E+i\delta} \right)
\langle \Phi_0|\sum_i z_i|\Phi_n \rangle
\langle \Phi_n|\sum_i (\vec r \times \vec \nabla)_z |\Phi_0 \rangle.
\end{eqnarray}
${\cal R}(E)$ of eq.(7) is the sum over three directions,
${\cal R} = {\cal R}_x + {\cal R}_y + {\cal R}_z$.

In the time-dependent local density approximation, the time-dependent
wave function $\Psi(t)$ is approximated by a single Slater determinant.
We prepare the initial single electron orbitals as $\psi_i(0)=
\exp[ikz]\phi_i$ where $\phi_i$ is the static Kohn-Sham orbitals
in the ground state. The $\psi_i(t)$ follows the time-dependent
Kohn-Sham equation,
\be
i\hbar \frac{\partial}{\partial t} \psi_i(t)
=
\left\{ -\frac{\hbar^2}{2m}\nabla^2 + \sum_a V_{ion}(\vec r - \vec R_a)
+e^2 \int d\vec r' \frac{\rho(\vec r',t)}{|\vec r - \vec r'|}
+\mu_{xc}(\rho(\vec r,t))
\right\}
\psi_i(t),
\ee
where $V_{ion}$ is the electron-ion potential and $\mu_{xc}$ is the
exchange-correlation potential.
The time-dependent density is given by
$\rho(\vec r,t)=\sum_i |\psi(\vec r,t)|^2$.
The time-dependent dipole moment may be evaluated as
$z(t) = \sum_i \langle \psi_i |z| \psi_i \rangle$ and the
similar expression for $L_z(t)$.
The strength functions are then evaluated with eqs.(12) and (14).

\subsection{Sum rules}

According to the TRK sum rule, the integral of
$S(E)$ is equal to
the number of active electrons $N$. This sum rule is respected by the
TDLDA.  It also appears in the short time 
behavior of $z(t)$ as
\be
z(t) = N \frac{\hbar k}{m}t ~~~~~~~(t ~small)
\ee
For the rotational strength, we define energy-weighted sums as
\be
R^{(n)} = \int_0^{\infty} dE E^n R(E).
\ee
It is known that $R^{(n)}$ for $n \le 4$ vanishes identically in
the exact dynamics \cite{ha80,ha77,ca77,kh92}. The vanishing of
$R^{(0)}$ in the time-dependent Hartree-Fock theory was first
noticed in  \cite{ha69}.
The short time behavior of $L(t)=L_x(t)+L_y(t)+L_z(t)$
is related to $R^{(n)}$ as
\be
\frac{e^2 \hbar}{2mc}L(t) = 2k \left\{
R^{(0)} 
- \frac{R^{(2)}}{2!} \left( \frac{t}{\hbar}\right)^2 
+ \frac{R^{(4)}}{4!} \left( \frac{t}{\hbar}\right)^4 
- \frac{R^{(6)}}{6!} \left( \frac{t}{\hbar}\right)^6 +...
\right\}
\ee
Here we note that $L(t)$ is an even function of $t$ as seen in eq.(13).
Since $R^{(n)}=0$ for $n \le 4$, we see that $L(t)$ behaves as
$t^6$ for small time. $L_i(t)(i=x,y,z)$ behave as $t^2$ at
small $t$ and the cancelation up to $t^4$ order occurs after summing
up three directions. In the TDLDA dynamics, we confirmed that at least 
$t^0$ and $t^2$ coefficients of $L(t)$, namely $R^{(0)}$ and $R^{(2)}$, 
vanish identically.

\subsection{Numerical detail}

The TDLDA uses the same Kohn-Sham Hamiltonian as is used in ordinary
static LDA calculations.  As is common in condensed matter theory,
we use pseudopotentials that include the effects of $K$-shell electrons
rather than include these electrons explicitly.  The pseudopotentials
are calculated by the prescriptions of ref. \cite{tr91} and \cite{kl82}.  
We employ the simple exchange-correlation term  proposed in ref. 
\cite{ce80,pe81}.  There are improved terms now in use\cite{pe96,le94}, 
but it was not deemed important for our application.

There are many numerical methods to solve the equations of TDLDA.  
Ours uses a Cartesian coordinate space mesh to represent the electron 
wave functions, and the time evolution is calculated directly
\cite{ya96,ya97,ya98}.  
There are only four important numerical parameters in this approach:  
the mesh spacing, the number of mesh points, the time step in the time 
integration algorithm, and the total length of time that is integrated.
We have previously found that the carbon molecules can be well treated
using a mesh spacing of 0.3~\AA \cite{ya97,ya98}. 
We find 0.25~\AA ~is necessary for
methyloxirane to represent accurately the orbitals around oxygen.
We take the spatial volume to be a sphere of radius 8~\AA ~both for 
methyloxirane and C$_{76}$ presented below. The total number of mesh 
points, defining the size of the vector space, is about 
$4\pi R^3/(3\Delta x^3) \sim 80,000 ( 140,000)$ 
for mesh size of 0.3~\AA (0.25~\AA).

The algorithm for the time evolution is quite stable as long as the time
step $\Delta t$ satisfies $\Delta t < \hbar / |H|$, where $|H|$ is the
maximum eigenvalue of the Hamiltonian.  This is mainly dependent on
the mesh size. For $\Delta x = 0.3$~\AA, we find that
$\Delta t = 0.002\hbar$/eV is adequate. We integrate the equation
of motion for a length of time $T=60\hbar$/eV for C$_{76}$
($50\hbar$/eV for methyloxirane). Then individual states can be
resolved if their energy separation satisfies $\Delta E > 2\pi \hbar/T
\sim 0.1 $eV.

Our numerical implementation, grid representation of the wave function
and the time-frequency Fourier transformation for the response calculation,
has several advantages over the usual approach using
basis functions centered at the ions. They include:\\
(1) The full spectrum of wide energy region may be calculated at once
and it respects sum rules. The non-locality of the pseudopotential 
may cause violation of the sum rule, but the effect is small in the 
present systems.\\ 
(2)Since the circular dichroism $R(E)$ and the optical 
rotatory power, real part of $\beta(E)$, are calculated as Fourier 
transformation of single function $L(t)$, the Kramers-Kronig relation 
is automatically satisfied.\\ 
(3)The gauge independence of the results is 
satisfied to high accuracy. Employing the commutation relation
$[H,\sum_i \vec r_i] = -\frac{\hbar^2}{m} \sum_i \vec \nabla_i$,
there is alternative expressions for optical transitions with gradient
operator instead of coordinate. For the rotational strength, for example,
\be
R_n = -\frac{e^2 \hbar^3}{2m^2 c E_{n0}}
\langle \Phi_0 |\sum_i \vec \nabla_i |\Phi_n \rangle \cdot
\langle \Phi_n |\sum_i {\vec r}_i \times {\vec \nabla}_i | \Phi_0 \rangle,
\ee
The strength function with this expression may be calculable with 
initial wave function $\psi_i(0)=\exp[id\nabla_z]\phi_i$ with small 
displacement parameter $d$. 
Since the grid representation of wave function does not have any
preference on the gauge condition, our method gives almost identical
results for the coordinate and gradient expressions of dipole matrix
elements.

\section{ R-methyloxirane}

The geometry of R-methyloxirane is shown in Fig.~1.  We use the same  
nuclear distances as in ref.~\cite{ca91}.  
We show in Fig.~2 the results of the static calculation for the 
orbital energies. We find a LUMO 6.0 eV above the HOMO, and 
a triplet of unoccupied states 0.5 eV higher. In our calculation
the lowest unoccupied orbitals have an diffuse, Rydberg-like character, 
$s$-wave for the lower and $p$-wave for the upper, as in previous 
calculations\cite{ca91,ra81}.  The HOMO is localized in the vicinity of 
the oxygen atom, and the measured absorption strength seen at 7.1 and 
7.7 eV is attributed to the excitation of a HOMO electron to the 
diffuse states. In the TDLDA, the excitation energy comes out close to 
the orbital difference energies, except for strongly collective states.  
Indeed we find in the TDLDA calculation that the excitations are within 
0.1 eV of the HOMO-LUMO energy and the energy difference for the next state
above the LUMO.  This is one eV less than the experimental values.  It
is known that the LDA energy functional that we use is subject to overbind
excitations close to the ionization threshold.  There are improved
energy functions that rectify this problem\cite{ca98}, but for this work 
we judged the error not important. 

The next property we examine is related to the electric dipole matrix
element, namely the oscillator strength $f$ associated with the transition.
The optical absorption strength is shown in Fig.~3.  The
total strength up to 100 eV excitation is f=22.4, which is 93\% of
the sum rule for the 24 active electrons.
Notice the lowest two peaks, centered at 6.0 and 6.5 eV.  
These are the states we are interested in. 
Their oscillator strengths are given in Table I.  We see that the states 
are both weak, less than a tenth of a unit.  The effect of the 
time-dependent treatment is to lower the strengths by
30-50\%.  This is the well-known screening effect associated with virtual
transitions of more deeply bound orbitals.  We find that the computed
transition strengths are within a factor of two of the measured ones.
Typically, the TDLDA does somewhat better than this, but most of the
experience has been with transitions carrying at least a tenth of a unit 
of oscillator strength. The original theoretical
calculation gave very poor transition strengths\cite{co83}, off by more than
an order of magnitude.  Unfortunately,
the more recent study\cite{ca91} did not include theoretical transition
strengths.

We numerically confirmed that our method gives almost identical results 
with coordinate and gradient expressions of dipole matrix elements,
as we noticed in the previous section. However, exceptionally, the oscillator 
strength of the very weak features discussed above suffer substantial 
dependence on the expression. With gradient formula for the transition 
matrix elements, strengths of both first and second transitions are larger
by about factor two than the coordinate expression. Since the gradient 
formula emphasizes high-momentum components more heavily,
we think the results with coordinate matrix elements may be more reliable 
for low transitions, and we quote them in Table I. 

We now turn to the chiroptical response.  Fig.~4 shows the short-time 
behavior of $L_x(t)$ and the sum of the three Cartesian components 
$L(t)=\sum_i L_i(t)$. An initial perturbation of $k=0.001$ \AA$^{-1}$ is employed.
To within numerical precision, $L_x(t)$ (solid line) grows with time as 
$t^2$, as discussed below eq.(18).  The same is true for
the other two components, $L_y$ and $L_z$. This shows that the
numerical algorithm respects the first sum rule.  The combined
response, $L(t)$ (dashed lines) shows an extreme cancelation
at short times, as required by the additional sum rules.
However our numerical accuracy does not allow us to determine the order 
of the power behavior.
The evolution of $L(t)$ for larger times is shown in Fig.~5.
Physically, the TDLDA breaks down at long times because of coupling to
other degrees of freedom.  A typical width associates with such couplings
is of the order of a tenth of an eV, implying that the responses damp
out on a time scale of $T\approx 10 \hbar$/eV.  We note that the
TDLDA algorithm itself is very stable, and allows us to integrate
to much larger times and obtain very sharp theoretical spectra.

We next show the Fourier transform of the chiroptical response.
The circular dichroism spectrum calculated with eq.~(14) is shown in
Fig.~6.
Here we have integrated $L$ to $T=50\hbar$/eV, including a filter function in
the integration to smooth the peaks.  One can see that the $s$- and
the $p$-transitions are clearly resolved, although the three $p$-transitions
are not resolved from each other (as is the experimental case).  
The $s$-transition has a negative circular
dichroism and the $p$-transition a positive one. Integrating over the
peaks, the strengths of the two peaks are -0.0014 and +0.0014 \AA$^3$-eV, 
respectively. The strengths are commonly quoted in cgs units; the conversion 
factor is 1 eV-\AA$^3$ = 1.609 $\times 10^{-36}$ erg-cm$^3$.  
The values in cgs units are given in Table I, compared to experiment 
and previous calculations.  We find the signs are correctly
given, but the values are somewhat too high, by a factor of 2 or 3.  The 
calculation of ref.\cite{ca91} gave a result within the experimental
range for the $p$-multiplet but too small (by a factor of 2) 
for the $s$-transition.
Thus we find that the TDLDA has a somewhat poorer accuracy in this case.

Next we consider the optical rotatory power. It could be calculated
as the real part of the Fourier transformation eq.(14). In practice, 
however, we found the calculation employing Kramers-Kronig
relation to the rotational strength function,
\be
{\rm Re}\beta(E) = \frac{2}{3} \hbar c
\int_0^{\infty} dE' \frac{R(E')}{{E'}^2 - E^2},
\ee
gives more accurate result especially at the energy below
the lowest transition. The measurement is available at sodium D-line,
2.1 eV, $[\alpha]_D = +14.65^{\circ}$ \cite{ra81} 
which gives $\beta=+0.0017$ \AA$^4$.
The calculated value at low energy is very sensitive to the
number of states included in the sum. Fig.6 shows the calculated
value as a function of a cutoff energy, upper bound in the integration
in eq.(20). The value taking only the contribution of the lowest
transition is -0.06. Including more states produces values that
fluctuate in sign and magnitude within that range. 
Including all states below 100 eV gives a cancelation by a factor
of 60 to yield a value
$\beta=-0.001$ \AA$^4$. This has the opposite sign but the same
order of magnitude as the measured $\beta$.  Clearly, to get
high relative accuracy with such a strong cancelation is more demanding
than our TDLDA can provide.

\section{C$_{76}$}

Remarkably, it is possible to separate the chiral partners of the
double-helical fullerene C$_{76}$ using stereospecific chemistry\cite{ha93}.
The molecule shows a huge optical rotatory power, 
$[\alpha]_D = -4000^\circ$, 
and a complex circular dichroism spectrum 
between 2 and 4 eV excitation \cite{ha93}.
There has been reported only semi-empirical quantum chemistry calculation
for the optical activity of this molecule \cite{or94}.

We first remark on the geometry of the molecule, which has a chiral $D_2$
symmetry\cite{et91}.  The accepted geometry
is depicted in ref.\cite{ha93}; it may be understood as follows.  We start
with C$_{60}$, in which all carbons on pentagons.  Group the pentagons into
triangles and divide the fullerene in half keeping two adjacent triangles
of pentagons intact in each half.  The ``peel" of six pentagons already has
a chiral geometry dependent on the relative orientation of its
two triangles of pentagons. The C$_{76}$ is constructed by adding 16 carbon 
atoms between the split halves of the C$_{60}$. The added carbon atoms lie 
entirely on hexagons which form a complete band around the fullerene. The 
inserted band has the geometry of an (8,2) chiral buckytube.
The result is then the chiral C$_{76}$.  Our calculations are performed
on a right-handed C$_{76}$, in the sense that the band of hexagons corresponds
to a right-handed buckytube. This is the same convention as used in 
ref. \cite{et91}, their Fig.~3d.  

The C$_{76}$ has
152 occupied spatial orbitals.  We show the orbitals near the Fermi level
in Fig.~8.  The HOMO-LUMO gap is only 0.9 eV, and there are many transition
in the optical region.  In Fig.~9 we show the optical absorption 
strength function for the range 0-50 eV.  Smoothing is made with the width of
0.2 eV in the Fourier transformation. A concentration of
strength is apparent at 6 eV excitation; there is a similar peak
in graphite and \C60 which is associated with $\pi-\pi^*$ transitions.
The strong, broad peak centered near 20 eV is associated with
$\sigma-\sigma^*$ transitions and is also present in \C60 \cite{ya98}.  
The feature at 13 eV is not present in \C60, however.  In the 
next figure, Fig.~10, we show a magnified view of the absorption 
at low energy.  We also compare the TDLDA strength with the
single-electron strength, smoothed also by convolution with a Breit-Wigner
function of width $\Gamma=0.2$~eV.  The TDLDA has a strong influence
on the strength distribution, decreasing the total strength in the
low energy region and concentrating in the 6 eV peak.  The experimental
absorption strength \cite{et91} (with arbitrary normalization) is shown as the 
dashed line. It agrees with the TDLDA quite well.

We next examine the circular dichroism spectrum.  Fig.~11 shows the rotatory
strength function between 0 and 50 eV.  Like the case of methyloxirane,
it is irregular without any large scale structures.  Its integral is zero 
to an accuracy of 0.001 eV-\AA$^3$.  The low energy 
region is shown in Fig.~12.  Here one sees qualitative similarities 
between theory and experiment \cite{ha93}. 
The theoretical sharp negative peak at 1.8 eV
corresponds to an experimental peak at 2.2 eV.  Shifting the higher
spectra by that amount (0.6 eV),  one sees a 
correspondence between the next positive
and negative excursions. We note that a similar shift in the 
excitation energy was also seen in the optical absorption of \C60 between 
the measurement and the TDLDA calculation \cite{ya98}.

Our theoretical optical rotatory power is plotted in Fig.~13.
The situation here is different
from the methyloxirane, in that rotatory power is large in a region 
where there are allowed transitions. The measured optical
rotatory power, $[\alpha]_D = -4000^{\circ}$ at 2.1 eV \cite{ha93}
corresponding to $\beta=-7.3$~\AA$^4$, is shown as the star.  It does
not agree with theory, but we should remember that the spectrum needs
to be shifted by 0.6 eV to reproduce the circular dichroism.  
Adjusting the theoretical spectrum upward by that amount, we find that
it is consistent in sign and order of magnitude with the measurement.
Since the optical rotatory power in the region of allowed
transitions changes rapidly as excitation energy, a measurement of the
energy dependence would be very desirable, and would allow a more
rigorous comparison with the theory.

\section{Concluding remarks}

We have presented an application of the time-dependent local density
approximation to the optical activities of chiral molecules.
Our method is based on the uniform grid representation of the wave
function, real-time solution of the time-dependent Kohn-Sham equation,
and the time-frequency Fourier transformation to obtain the response
functions. In this way we can calculate the optical absorption, 
circular dichroism, and the optical rotatory power over a wide energy 
region, respecting sum rules and Kramers-Kronig relation.

We applied our method to two molecules, methyloxirane and C$_{76}$.
For lowest two transitions of methyloxirane, the TDLDA reproduces 
the absorption and rotational strength in the accuracy within factor two. 
The qualitative feature of the circular dichroism spectrum of C$_{76}$ 
is also reproduced rather well. However, the optical rotatory power
is found to be a very sensitive function with strong cancelation.
Even though we obtained rotational strength of full spectral region,
it is still difficult to make quantitative prediction of optical
rotatory power at low energies in our present approach.

\section{Acknowledgment}
We thank S. Saito for providing us with coordinates of C$_{76}$.  We also
thank him, S. Berry, and B. Kahr for discussions.
This work is supported in part by 
the Department of Energy  under Grant DE-FG-06-90ER40561, and by the
Grant-in-Aid for Scientific Research from the Ministry of Education, 
Science and Culture (Japan), No. 09740236. Numerical calculations were
performed on the FACOM VPP-500 supercomputer in the institute for
Solid State Physics, University of Tokyo, and on the NEC sx4 supercomputer
in the research center for nuclear physics (RCNP), Osaka University.

\newpage
\begin{table}
\caption{Transitions in R-methyloxirane}
\begin{tabular}{cc|cc|cc|cc}
   Level&  &TDLDA         &    LDA(Free)    &  \multicolumn{2}{c|}{Other
theory}& \multicolumn{2}{c}{Experiment}\\
& & & & ref. \protect\cite{ca91}&ref. \protect\cite{co83}&ref.
\protect\cite{ca91}&ref. \protect\cite{co83}\\
\tableline
1 & $E$ (eV) & 6.0& 5.97& 6.25(7.12) &6.40& 7.07 & 7.12 \\
  & $f$ & 0.012 & 0.021 & & 0.0004$^*$ & & 0.025 \\
  & $R$ ($\times 10^{40}$ cgs)& -23.0& -10.4 & -6.43 & -2.66$^{*,**}$ & -12.56 & -11.8$^{**}$\\
\tableline
2-4 & $\overline E$ (eV)& 6.5 & 6.55 & 6.95 &7.3 &7.70& 7.75 \\
 &$\Sigma f$& 0.044& 0.069& & 0.0012$^{*}$ & & 0.062  \\
 &$\Sigma R$ ($\times 10^{40}$ cgs)& 23. & 29.9 & 7.93 & 2.24$^{*,**}$ & 6.98 & 10.8$^{**}$ \\
\end{tabular}
$^{*}$ Calculation with coordinate expression of dipole moment. \newline
$^{**}$ Negative of value for S-methyloxirane.
\end{table}

\newpage
\section*{Figure Captions}
{Fig.~1.~View of R-methyloxirane with hydrogen on the chiral
carbon in the back (and not seen).  The chirality is R because
the three other bonds are arrange clockwise in the sequence O, CH$_3$,
CH$_2$.}

Fig.~2~Static LDA orbitals in methyloxirane.

Fig.~3~Optical absorption strength of methyloxirane in the energy
region 0-50 eV.

Fig.~4~Short-time chiroptical response of R-methyloxirane.  
Solid line is $L_x(t)$, dashed line is $\sum_i L_i(t)$.

Fig.~5~Chiroptical response of R-methyloxirane $L(t)$ for longer times.

Fig.~6~$R$ in R-methyloxirane in the interval 0-20 eV. 

Fig.~7 Optical rotatory power Re$\beta$ at $E=2.1$ eV as a function of 
cutoff energy $E_{max}$ in the integration of eq.~(20).

Fig.~8 Static LDA orbitals in C$_{76}$ near the Fermi level.

Fig.~9 Optical absorption spectrum of C$_{76}$ in the range 0-50 eV.

Fig.~10~ Optical absorption spectrum of C$_{76}$ in the range 0-8 eV.  Dotted 
line is the single-electron strength, solid line the TDLDA, and
dashed line experiment\protect\cite{et91}.

Fig.~11~ Circular dichroism spectrum of C$_{76}$.

Fig.~12~ Circular dichroism spectrum of C$_{76}$ comparing theory
(solid line) and experiment (dashed line).  The experimental data
is from ref. \protect\cite{ha93} and is with arbitary normalization.

Fig.~13~ Optical rotatory power of C$_{76}$, given as Re$\beta$ in unit
of \AA$^4$. The cross is the measured value from $[\alpha]_D$.

\newpage

\begin{figure}
  \begin{center}
    \leavevmode
    \parbox{0.9\textwidth}
           {\psfig{file=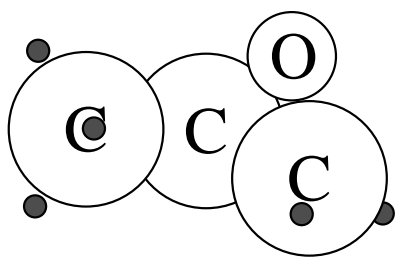,width=0.9\textwidth}}
  \end{center}
\caption{
View of R-methyloxirane with hydrogen on the chiral
carbon in the back (and not seen).  The chirality is R because
the three other bonds are arrange clockwise in the sequence O, 
CH$_2$O,CH$_3$.}
\label{geometry}
\end{figure}

\begin{figure}
  \begin{center}
    \leavevmode
    \parbox{0.9\textwidth}
           {\psfig{file=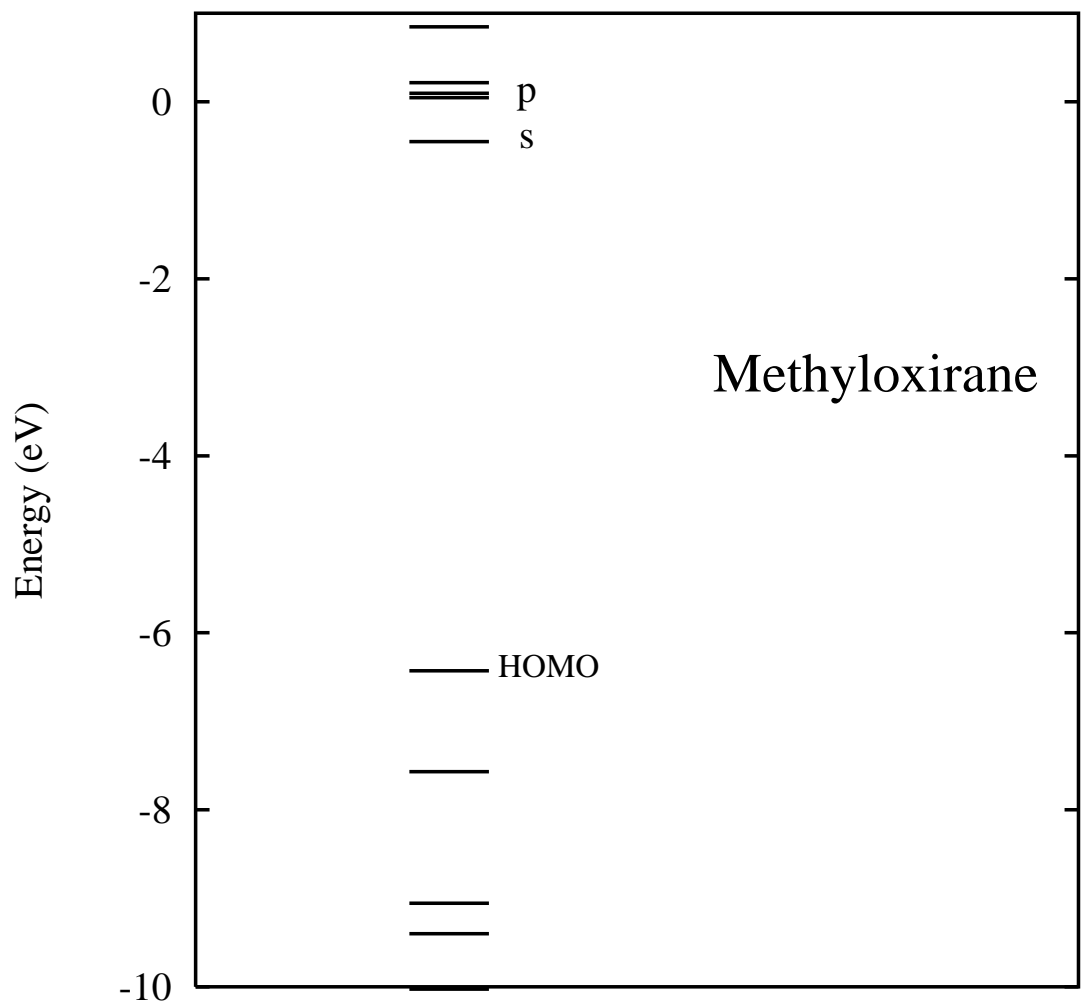,width=0.9\textwidth}}
  \end{center}
\caption{Static LDA orbitals in methyloxirane.}
\label{c3h6o-levels}
\end{figure}

\begin{figure}
  \begin{center}
    \leavevmode
    \parbox{0.9\textwidth}
           {\psfig{file=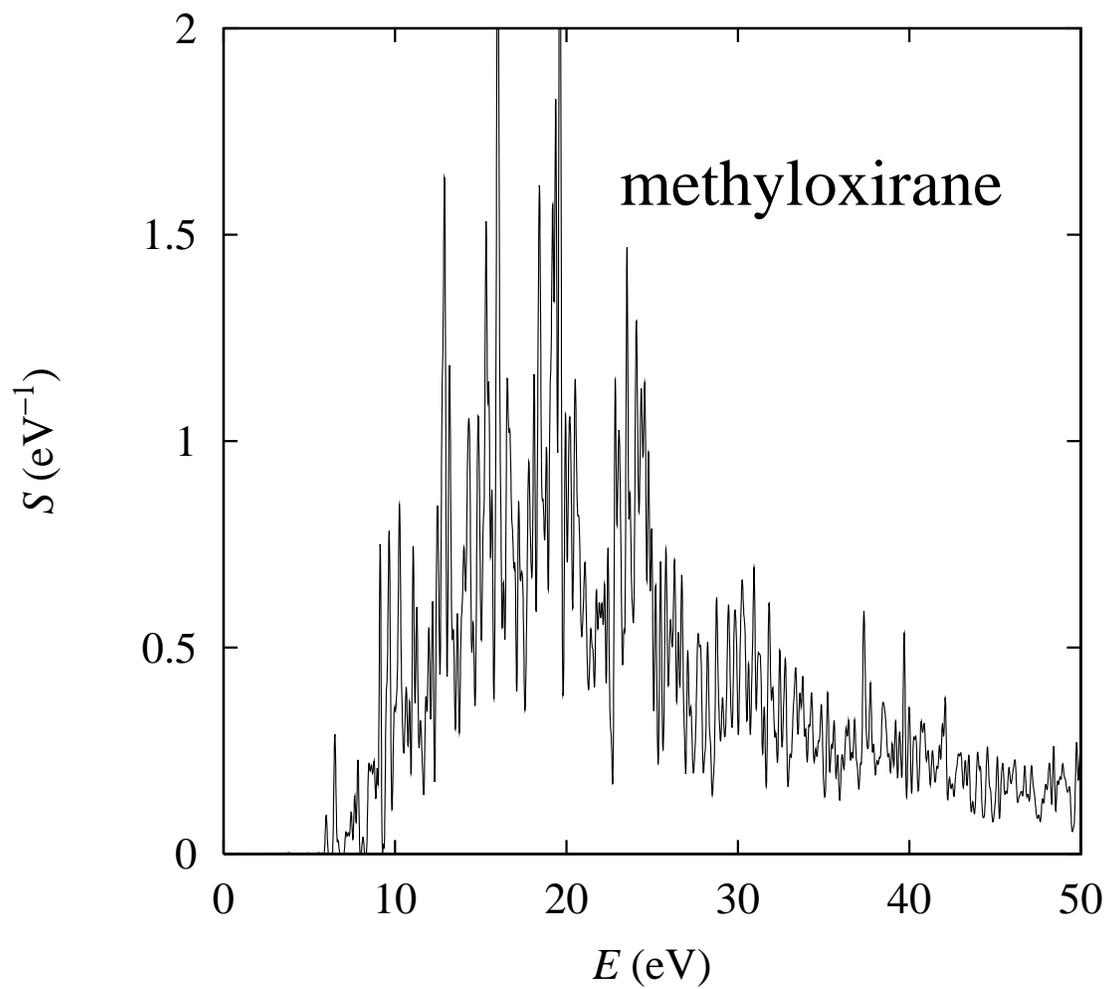,width=0.9\textwidth}}
  \end{center}
\caption{
Optical absorption strength of methyloxirane in the energy
region 0-50 eV.
}
\label{c3h6o-f}
\end{figure}

\begin{figure}
  \begin{center}
    \leavevmode
    \parbox{0.9\textwidth}
           {\psfig{file=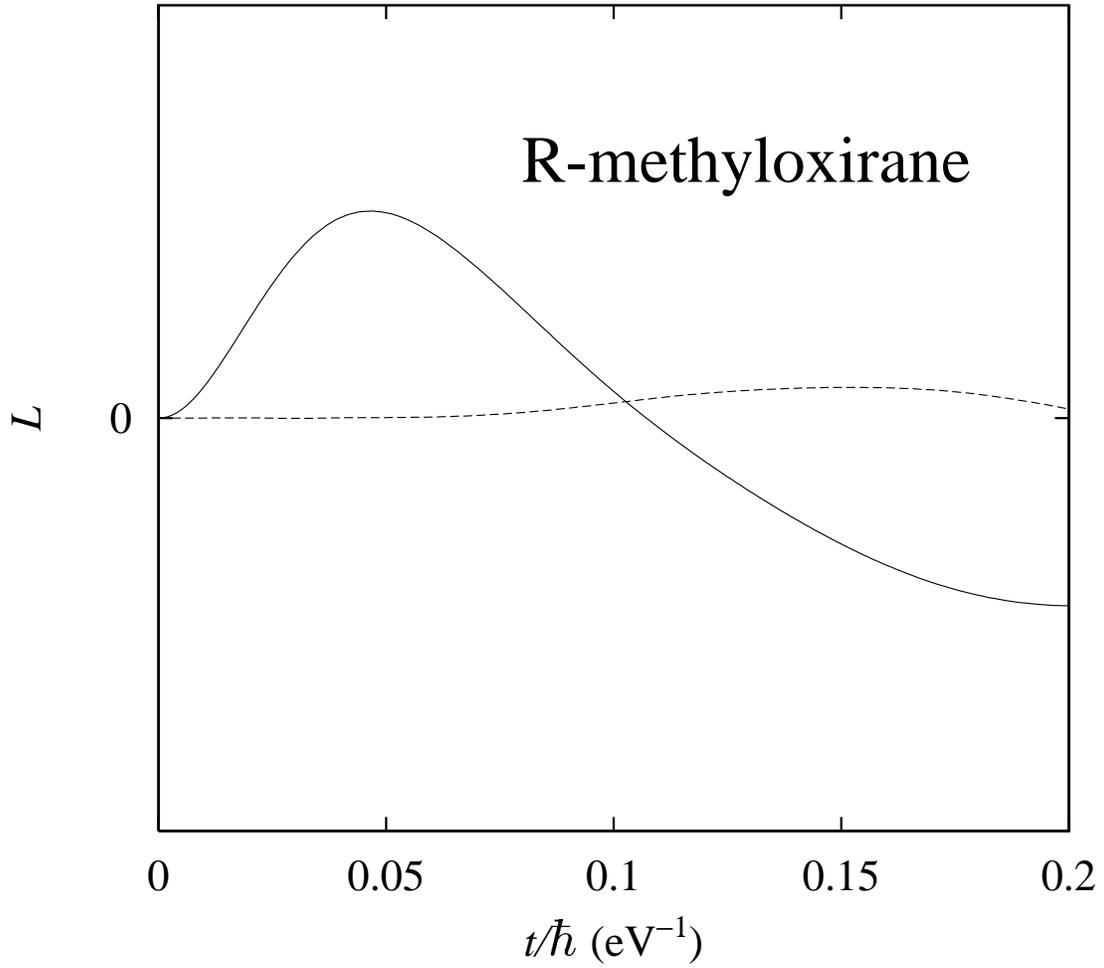,width=0.9\textwidth}}
  \end{center}
\caption{Short-time chiroptical response of R-methyloxirane.  
Solid line is $L_x(t)$, dashed line is $\sum_i L_i(t)$, in units
of \AA.
}
\label{time}
\end{figure}

\begin{figure}
  \begin{center}
    \leavevmode
    \parbox{0.9\textwidth}
           {\psfig{file=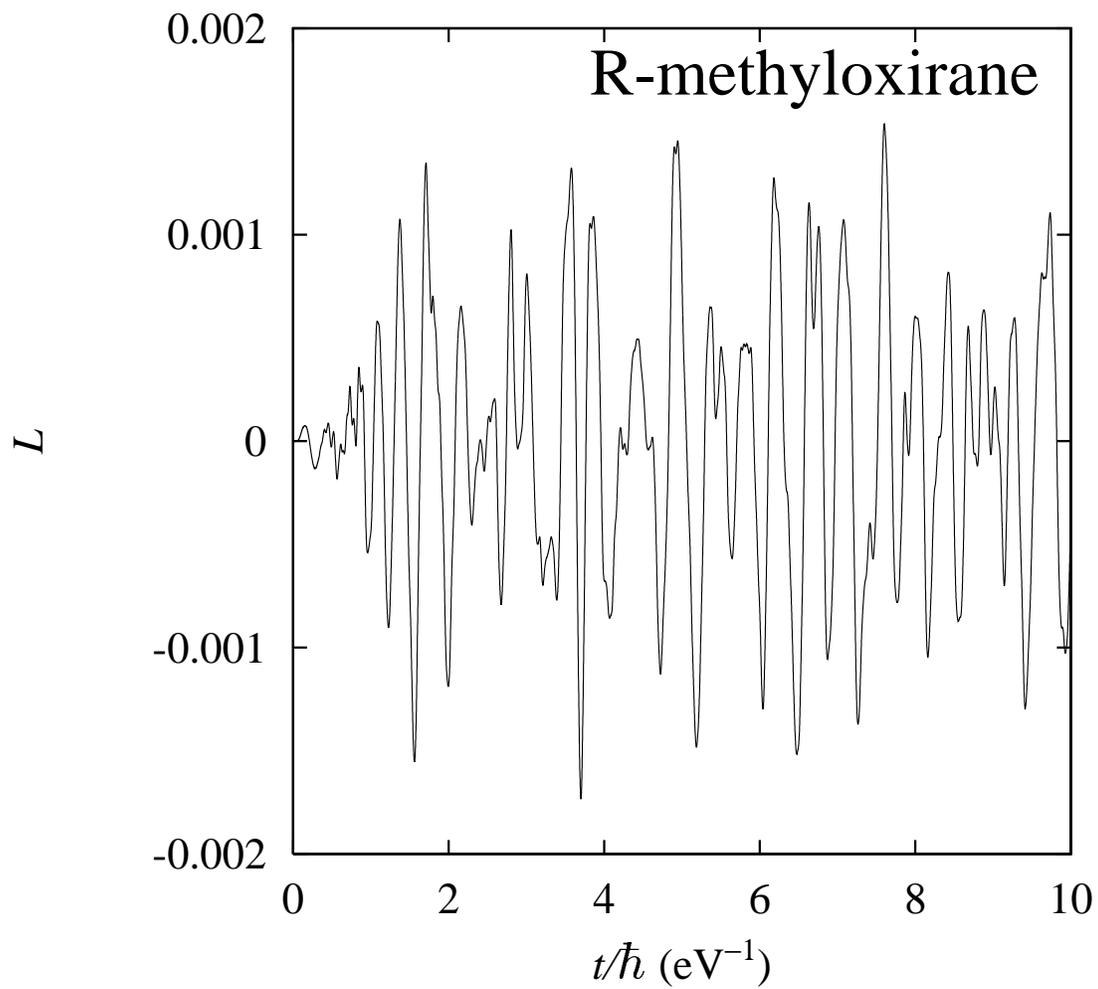,width=0.9\textwidth}}
  \end{center}
\caption{
Chiroptical response of R-methyloxirane $L(t)$ for longer times.
}
\label{time2}
\end{figure}

\begin{figure}
  \begin{center}
    \leavevmode
    \parbox{0.9\textwidth}
           {\psfig{file=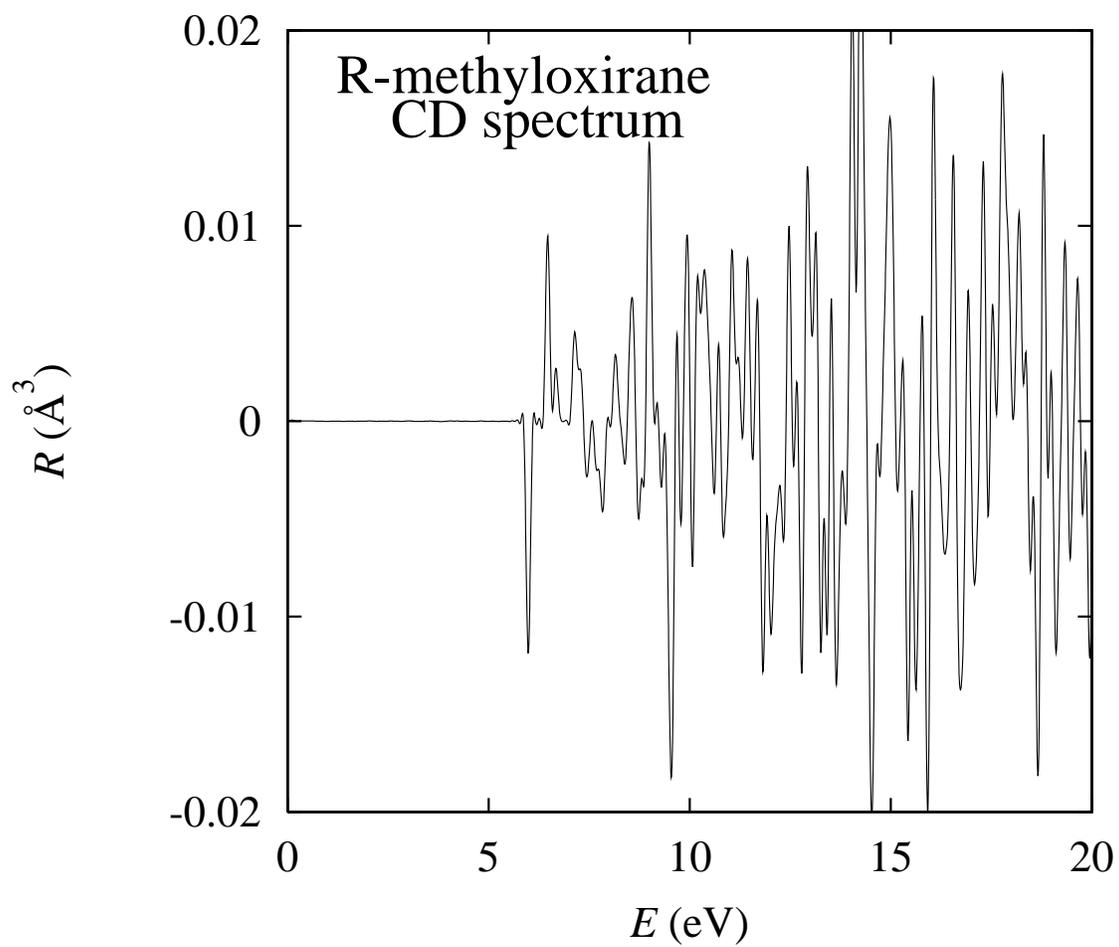,width=0.9\textwidth}}
  \end{center}
\caption{
$R$ in R-methyloxirane in the interval 0-20 eV. 
}
\label{cd-c3h6o}
\end{figure}

\begin{figure}
  \begin{center}
    \leavevmode
    \parbox{0.9\textwidth}
           {\psfig{file=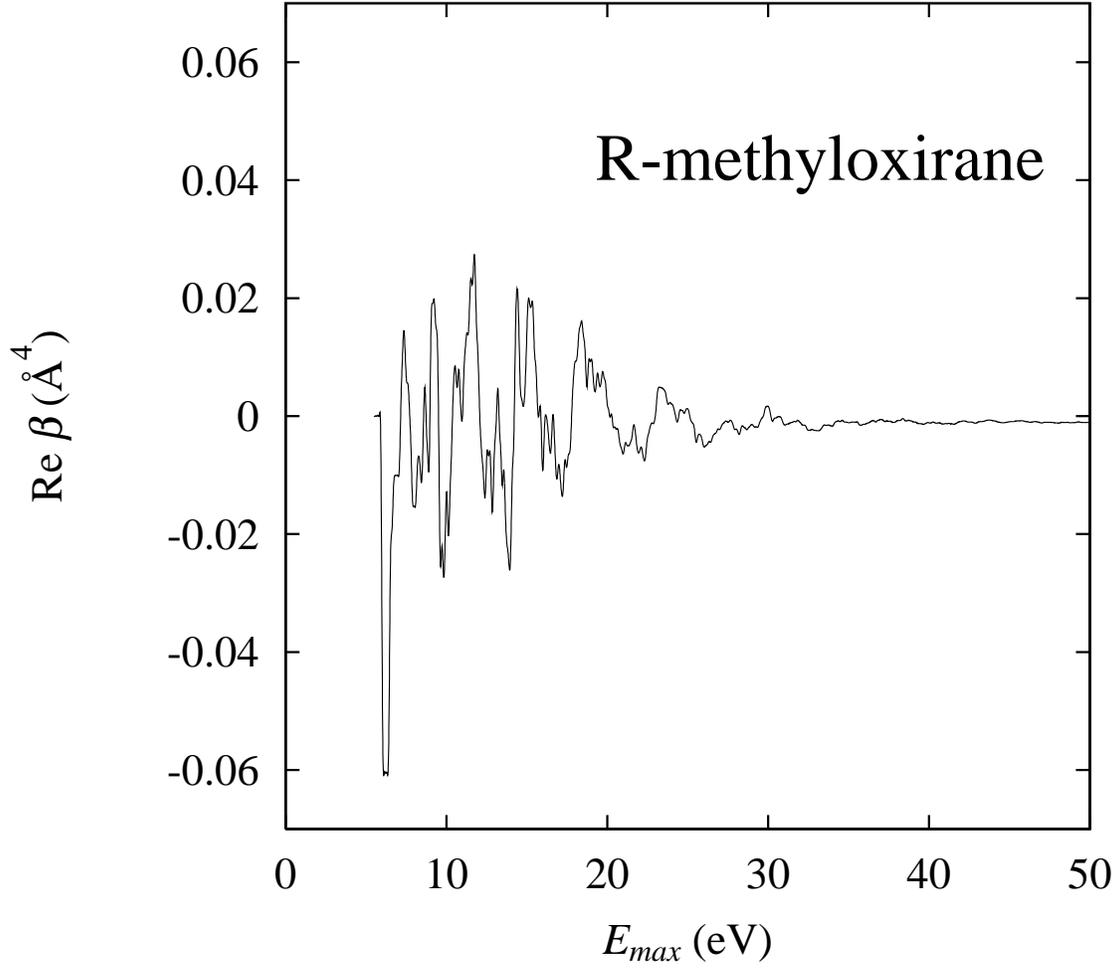,width=0.9\textwidth}}
  \end{center}
\caption{
Optical rotatory power Re$\beta$ at $E=2.1$ eV as a function of 
cutoff energy $E_{max}$ in the integration of eq.~(20).
}
\label{cd-c3h6oR}
\end{figure}

\begin{figure}
  \begin{center}
    \leavevmode
    \parbox{0.9\textwidth}
           {\psfig{file=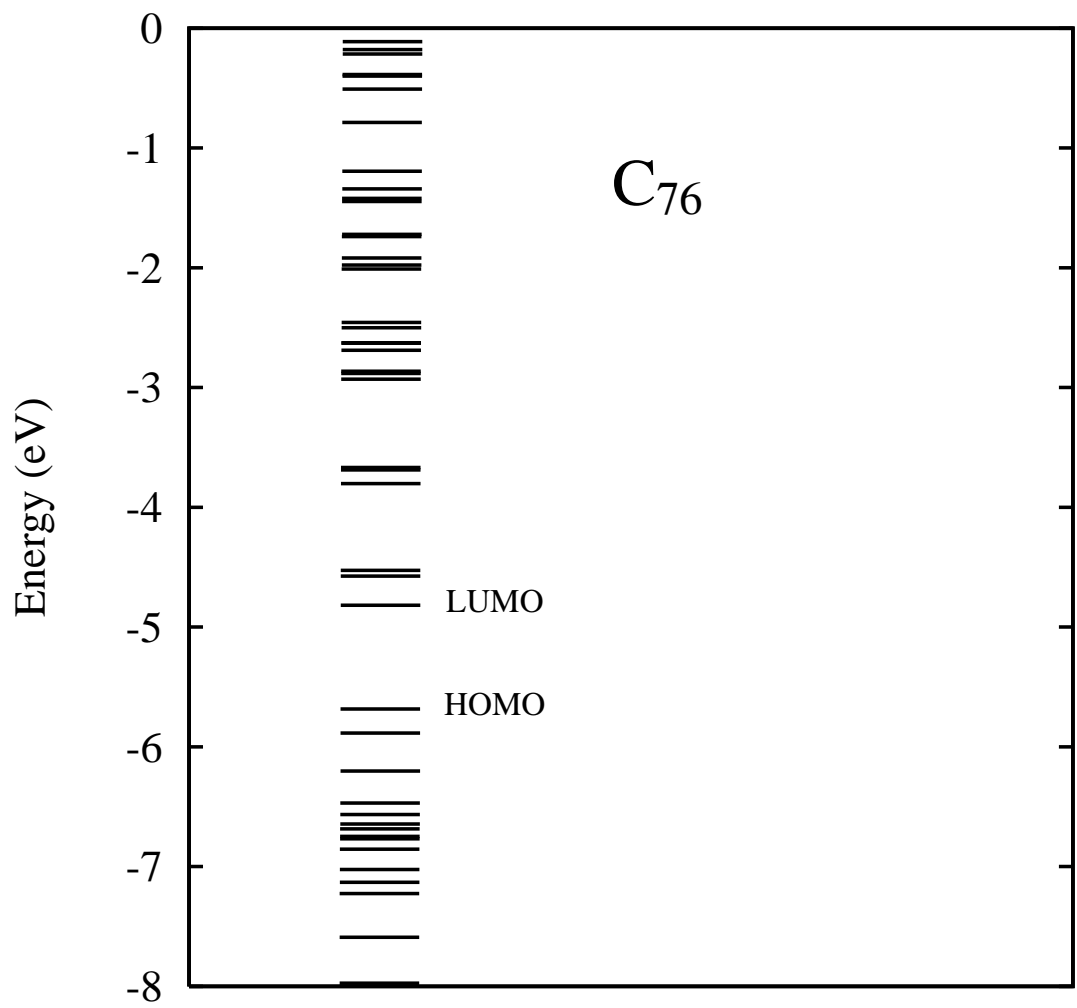,width=0.9\textwidth}}
  \end{center}
\caption{
Static LDA orbitals in C$_{76}$ near the Fermi level.
}
\label{c76-levels}
\end{figure}
\begin{figure}
  \begin{center}
    \leavevmode
    \parbox{0.9\textwidth}
           {\psfig{file=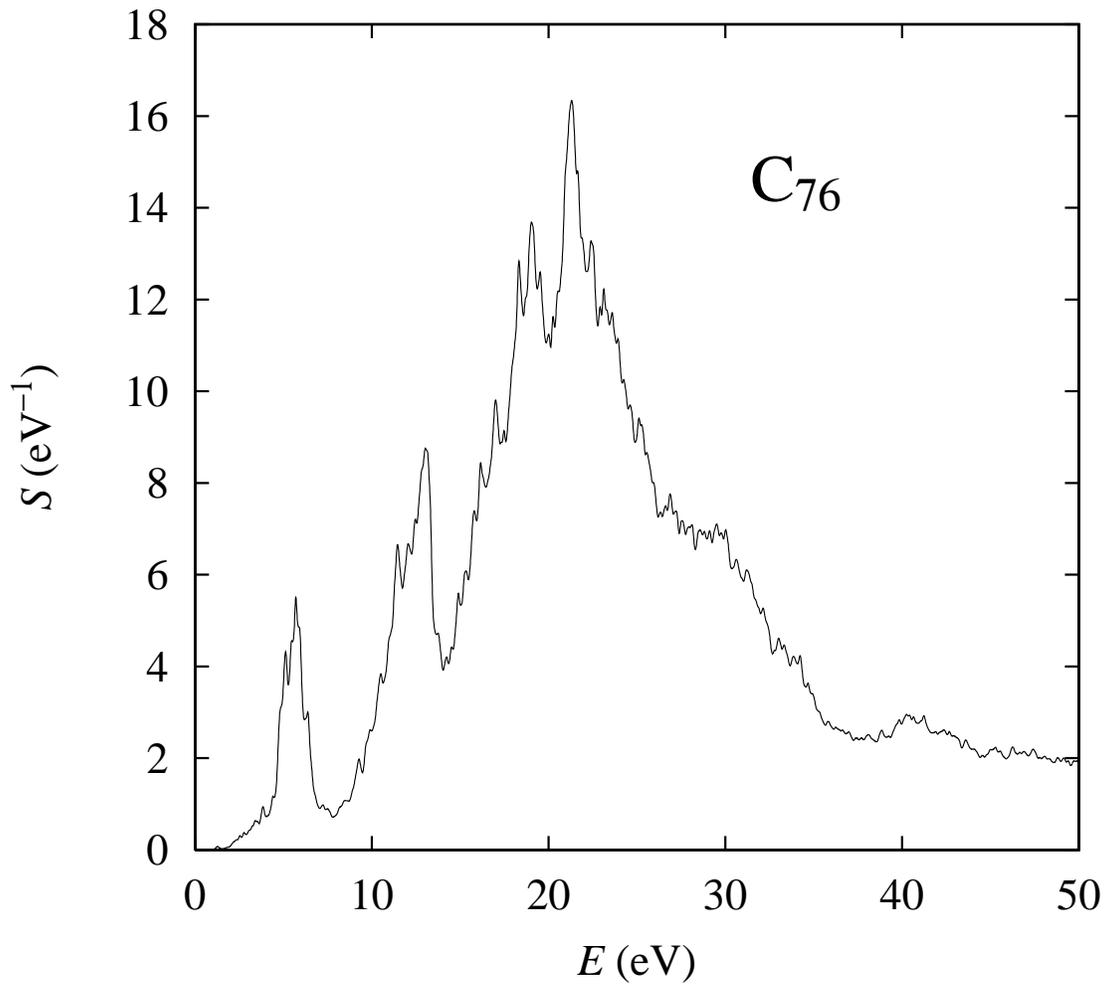,width=0.9\textwidth}}
  \end{center}
\caption{
Optical absorption spectrum of C$_{76}$ in the range 0-50 eV.
}
\label{c76-s}
\end{figure}
\begin{figure}
  \begin{center}
    \leavevmode
    \parbox{0.9\textwidth}
           {\psfig{file=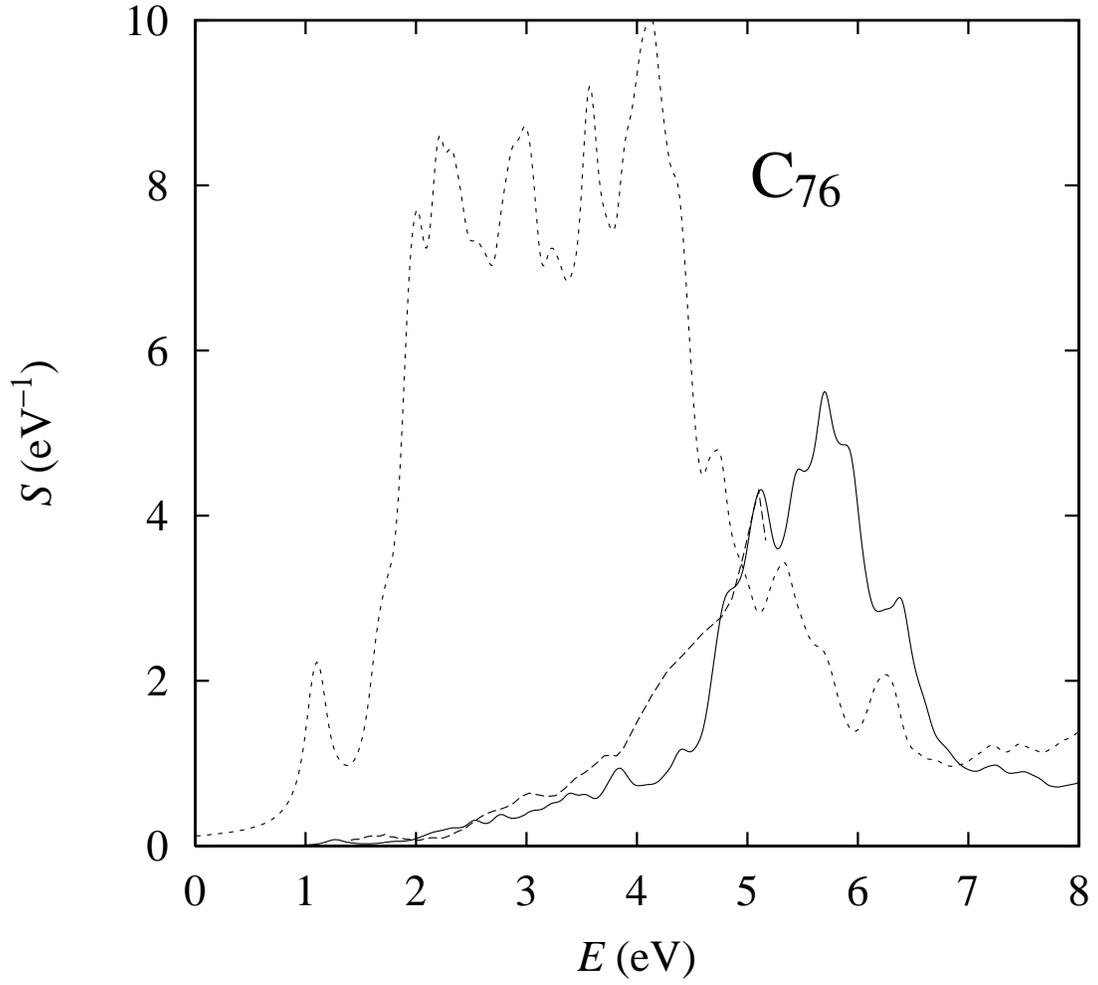,width=0.9\textwidth}}
  \end{center}
\caption{
Optical absorption spectrum of C$_{76}$ in the range 0-8 eV.  Dotted 
line is the single-electron strength, solid line the TDLDA, and
dashed line experiment\protect\cite{et91}.
}
\label{c76-sp}
\end{figure}

\begin{figure}
  \begin{center}
    \leavevmode
    \parbox{0.9\textwidth}
           {\psfig{file=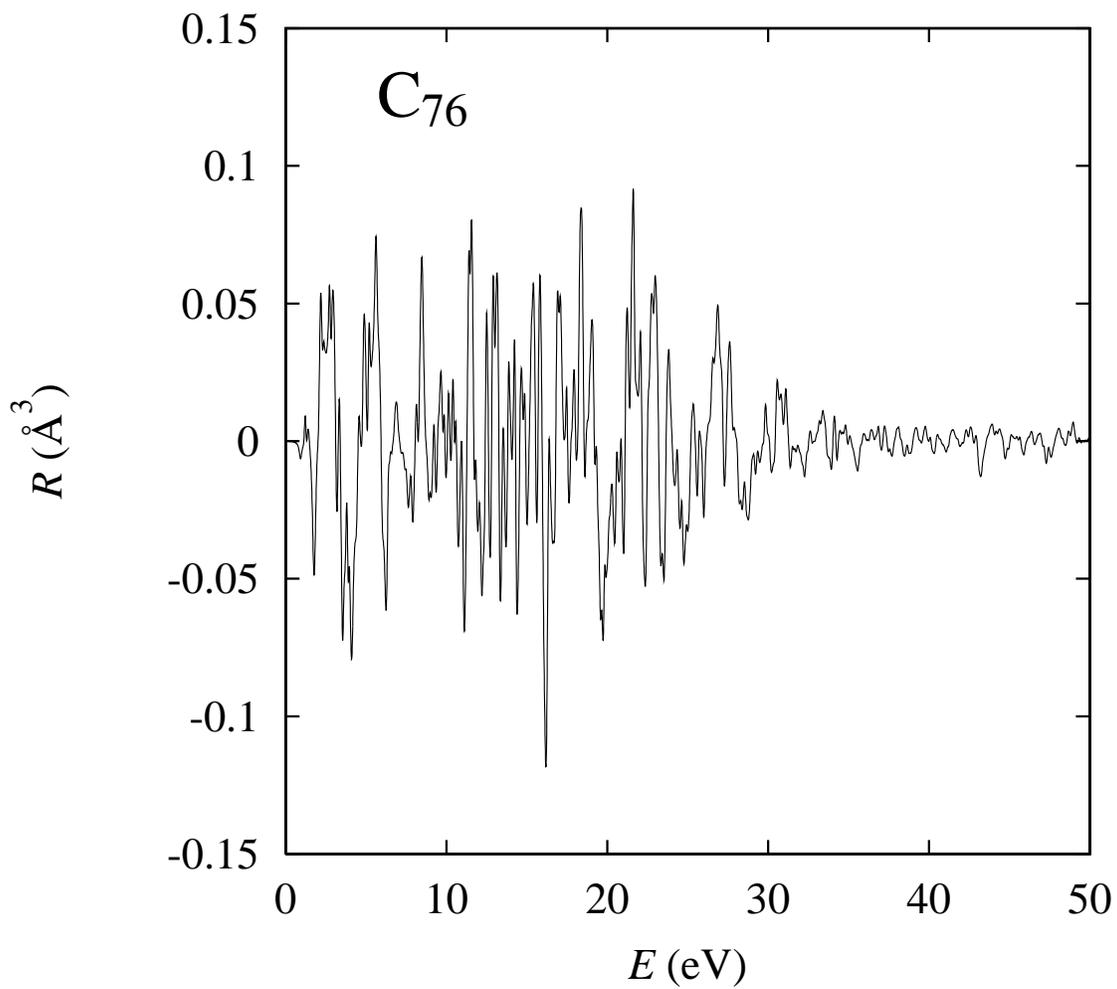,width=0.9\textwidth}}
  \end{center}
\caption{
Circular dichroism spectrum of C$_{76}$.
}
\label{cd-c76-whole}
\end{figure}

\begin{figure}
  \begin{center}
    \leavevmode
    \parbox{0.9\textwidth}
           {\psfig{file=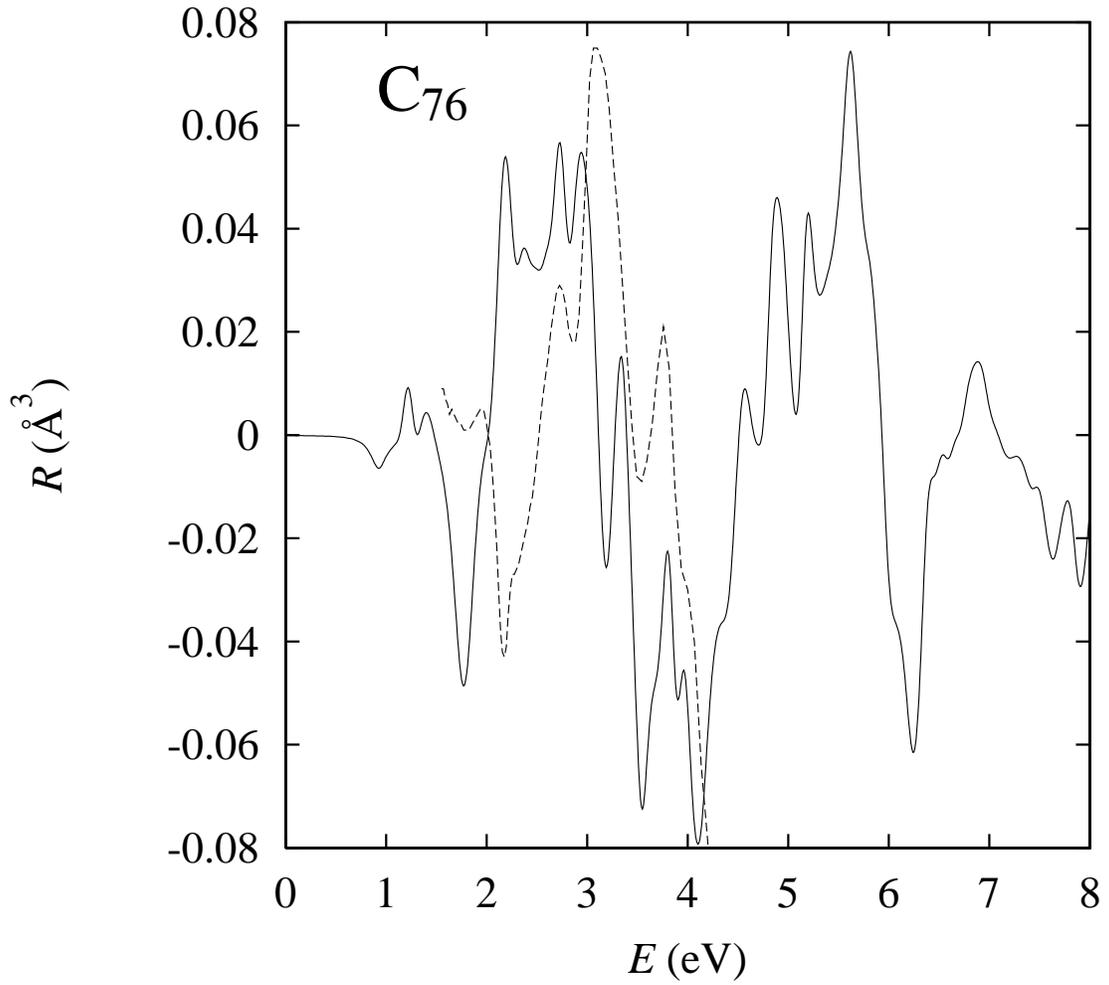,width=0.9\textwidth}}
  \end{center}
\caption{
Circular dichroism spectrum of C$_{76}$ comparing theory
(solid line) and experiment (dashed line).  The experimental data
is from ref. \protect\cite{ha93} and is with arbitary normalization.
}
\label{cd-c76-visible}
\end{figure}

\begin{figure}
  \begin{center}
    \leavevmode
    \parbox{0.9\textwidth}
           {\psfig{file=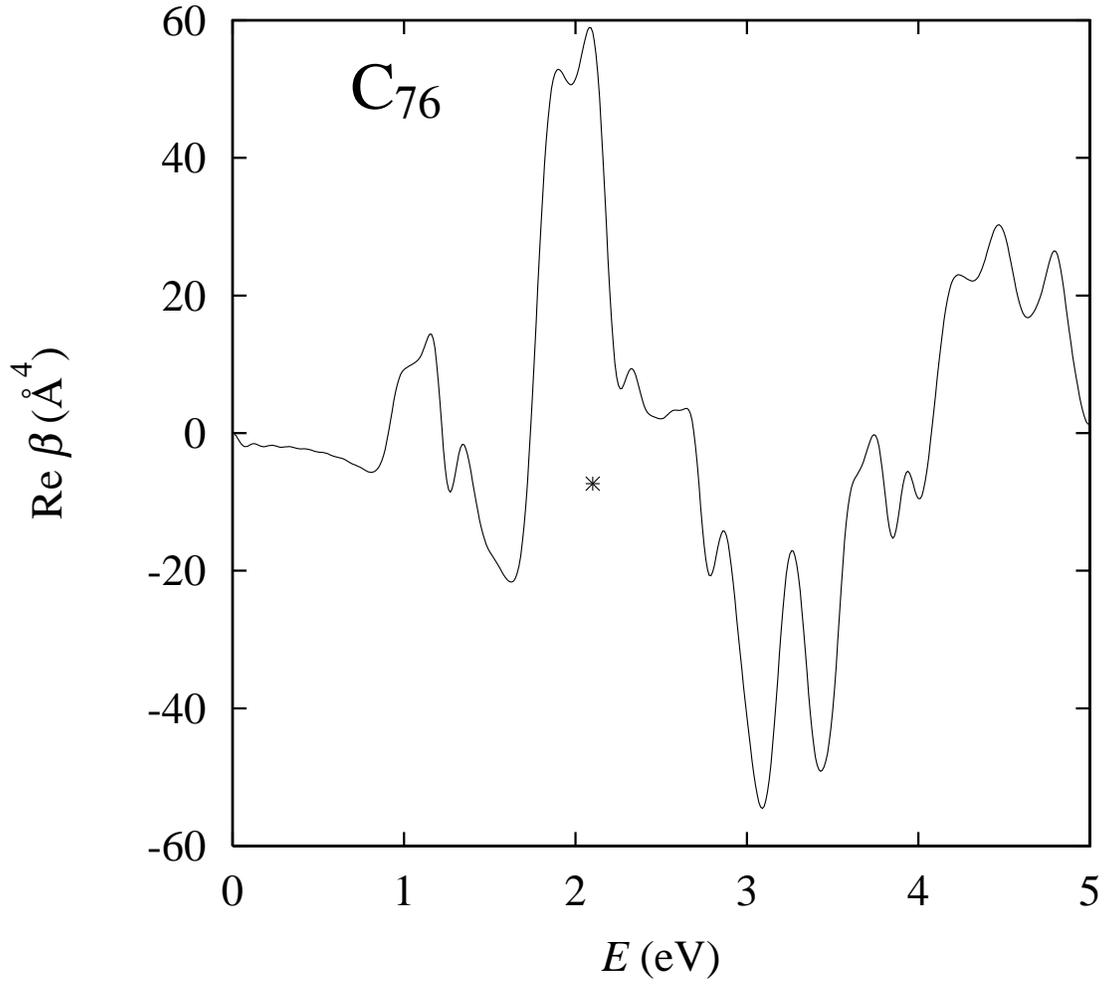,width=0.9\textwidth}}
  \end{center}
\caption{
Fig.~13~ Optical rotatory power of C$_{76}$, given as Re$\beta$ in unit
of \AA$^4$. The cross is the measured value from $[\alpha]_D$.
}
\label{cd-c76-rotatory}
\end{figure}

\end{document}